\documentclass[twocolumn,secnumarabic,amssymb, nobibnotes, aps, prd]{revtex4-1}
\usepackage{amsmath}
\usepackage{graphicx}
\usepackage{multirow}
\RequirePackage[pagewise,mathlines]{lineno}

\usepackage{float}
\begin{document}
\title{Photoproduction of charged final states in ultra-peripheral collisions and electroproduction at an electron-ion collider}%
	\author{Spencer. R. Klein}\email{srklein@lbl.gov}
\affiliation{Nuclear Science Division, Lawrence Berkeley National Laboratory, Berkeley, CA 94720 USA}
\author{Ya-Ping Xie}\email{xieyp@lbl.gov}
\affiliation{Institute of Modern Physics, Chinese Academy of
	Sciences, Lanzhou 730000, China, and\\
Nuclear Science Division, Lawrence Berkeley National Laboratory, Berkeley, CA 94720 USA}

\date{\today}%
\begin{abstract}

Ultra-peripheral collisions (UPCs) of relativistic ions are an important tool for studying photoproduction at high energies.  Vector meson photoproduction is an important tool for nuclear structure measurements and other applications.   A future electron-ion collider (EIC) will allow additional studies, using virtual photons with a wide range of $Q^2$.  We propose a significant expansion of the UPC and EIC photoproduction physics programs to include charged final states which may be produced via Reggeon exchange.   We consider two examples: $a_2^+(1320)$, which is a conventional $q\overline q$ meson, and the exotic $Z_c^+(4430)$ state (modeled here as a tetraquark).  The $Z_c^+(4430)$ cross-section depends on its internal structure, so photoproduction can test whether the $Z_c^+(4430)$ is a tetraquark or other exotic object.  We calculate the rates and kinematic distributions for $\gamma p\rightarrow X^+n$ in $pA$ UPCs and $ep$ collisions at an EIC and in UPCs.  The rates are large enough for detailed studies of these final states.   Because the cross-section for Reggeon exchange is largest near threshold, the final state rapidity distribution depends on the beam energies.  At high-energy colliders like the proposed LHeC or $pA$ collisions at the LHC, the final states are produced at far forward rapidities.   For lower energy colliders, the systems are produced closer to mid-rapidity, within reach of central detectors. 
\end{abstract}
\maketitle

{\it Introduction.}$-$Ultra-peripheral collisions (UPCs) are currently our main tool for studying photoproduction at high energies, above the reach of fixed-target experiments. UPCs are studied at both the Relativistic Heavy Ion Collider (RHIC) and the Large Hadron Collider (LHC) \cite{Baur:2001jj,Bertulani:2005ru,Baltz:2007kq}.  They have been used to probe both two-photon processes, such as lepton pair production and light-by-light scattering, and photonuclear interactions, such as vector meson and dijet production in heavy ions collisions \cite{Klein:2017vua}.  Photoproduction has been studied for  a variety of vector mesons, including the $\rho$, $\rho'$, $\omega$, $J/\psi$, $\psi'$ and $\Upsilon$, on both proton and ion targets.  At the LHC, these reactions occur predominantly via photon-Pomeron fusion, while, for the $\rho$ at RHIC, photon-Reggeon fusion also plays a role. A future electron-ion collider will allow studies of photoproduction using photons with significant $Q^2$ \cite{Accardi:2012qut}, and, by virtue of its high luminosity, expand the range of mesons that can be studied.   At still lower energies, hadrons are produced via the decay  of photo-excited baryon resonances.  These processes are relevant very near threshold, but we do not consider them here. 

In this Letter, we discuss a new class of particles that can be studied with UPC and EIC photoproduction:  electrically charged final states which are produced by the exchange of a charged Reggeon \cite{Regge:1959mz,Regge:1960zc,Leith:1977pg}. The Reggeons represent meson trajectories - the sum of meson exchange contributions.  Different Regge trajectories allow the exchange of different spins, $J$, with either natural ($P=(-1)^J$) or unnatural ($P=(-1)^{J+1}$) parity.  Because of this diversity of Reggeon trajectories, a wide range of final states can be produced.  We consider two examples: the $a_2^+(1320)$, which is a conventional $q\overline q$ meson, which is produced via $\gamma p\rightarrow a_2^+(1320) n$, the manifestly exotic final state $Z_c^+(4430)$, which we model  as a $u\overline d c\overline c$ tetraquark, produced by $\gamma p\to Z^+_c(4430)n$.  The $a_2^+(1320)$ is an attractive experimental target, since 70\% of it's decays involve 3 pions, leading to a good fraction of all-charged-particle final states.  Photoproduction is an attractive reaction to search for exotic hadrons, since, compared to hadroproduction, production of exotics is enhanced \cite{Szczepaniak:2001qz}.

The $Z_c^+(4430)$ was discovered by the Belle collaboration \cite{Choi:2007wga}.  It includes a $c\overline c$ pair, but is also charged, so it cannot be a conventional $c\overline c$ meson. Different theoretical interpretations have treated it as a tetraquark ($u\overline d c\overline c$), a hadronic molecule or a hadro-quarkonium state~\cite{Esposito:2014rxa}, with the tetraquark explanation attracting the most attention.  A measurement of the photoproduction cross-section would help differentiate between these models.   Several authors have studied the property of $Z_c^+(4430)$ in different reactions by using the effective Lagrangian method~\cite{Liu:2008qx,Ke:2008kf,Galata:2011bi,Wang:2015lwa}.    The $Z_c^+(4430)$ decays to $J/\psi \pi^+$ or $\psi(2S)\pi^+$, so is also experimentally tractable, albeit with a lower branching ratio to easily reconstructible states like $e^+e^-\pi^+$ or $\mu^+\mu^-\pi^+$.  A similar approach could be applied to other conventional and exotic hadrons. For example, the $Z_c^+ (3900)$ is lighter than the $Z_c^+(4430)$ , so if it is a similar class of hadron, is should have a higher production rate than the $Z_c^+(4430)$.

Because these interactions involve charge exchange, we consider only proton targets: $pA$ UPCs where the photon comes from the heavy nucleus, and $ep$ collisions at an EIC. Nuclear targets are interesting, but there are theoretical uncertainties in extending these photoproduction calculations to nuclear targets.  Photoproduction measurements are an important test of exotic hadron structure; heavy objects like the $Z_c^+(4430)$ are beyond the range of fixed-target photon beams, so UPCs and EICs are a unique probe of these heavy exotic states.   Reggeon exchange reactions also occur in $pp$ UPCs, but the cross-sections are smaller and the backgrounds larger, so we will not consider them here. 

{\it Cross-section calculations}$-$We extend two existing Monte Carlo codes to model the photoproduction of these objects.  For UPCs, we use STARlight \cite{Klein:2016yzr} which is widely used for photoproduction.  For $ep$ collisions, we use eSTARlight, which simulates vector meson photoproduction and electroproduction  \cite{Lomnitz:2018juf,Lomnitz:2018axr}.   Both codes make use of parameterized data.  For the $a_2^+ (1320)$, we added a Reggeon-inspired parameterization of cross-section data from fixed-target photoproduction experiments, while for the $Z_c^+(4430)$, we used a theoretical prediction which assumes that the $Z_c^+$ is a tetraquark with a $J^P=1^+$ states.  The $Z_c^+(4430)$ photoproduction cross-section is calculated in a Reggeon exchange model \cite{Galata:2011bi}.  The photoproduction cross-section is  sensitive to the $Z_c^+(4430)$ spin and parity;  $J^P=1^-$ leads to a 40\% larger cross-section, while  $J^P=0^-$ leads to a cross-section about three times larger at the peak, with the ratio increasing at larger photon energies.

We consider the six accelerator configurations shown in Tab. \ref{tab:parameters}: $pAu$ collisions at RHIC, $pPb$ collisions at the LHC, and $ep$ collisions at four proposed accelerators: the U.S. based eRHIC and JLEIC, CERN's proposed LHeC and the proposed Chinese EIC, EicC.   

\begin{table}
\begin{center}
\begin{tabular}{|c |c |c |c |c |}
\hline
\hline
Accelerator& AB & $e/p$ Energy & $p$ Energy & integrated luminosity\\
\hline
eRHIC \cite{Montag:2017}&  $ep$  &  18 GeV & 275 GeV & 10 $\mathrm{ fb}^{-1}$\\
JLEIC\cite{Morozov:2017}  & $ep$   & 10 GeV  & 100 GeV &10 $\mathrm{ fb}^{-1}$\\
LHeC\cite{AbelleiraFernandez:2012cc} & $ep$     & 60 GeV  & 7000 GeV &10 $\mathrm{ fb}^{-1}$\\
EicC \cite{Chen:2018wyz} & $ep$     & 3.5 GeV   & 20 GeV &10 $\mathrm{ fb}^{-1}$\\
RHIC\cite{Tanabashi:2018oca}  &$pAu$  & 100 GeV  & 100 GeV &4.5 $\mathrm{ pb}^{-1}$\\
LHC  \cite{Citron:2018lsq} & $pPb$ &7000 GeV & 2778 GeV & 2 $\mathrm{ pb}^{-1}$\\
\hline
\hline
\end{tabular}
\caption{Parameters for the accelerators. For eRHIC and JLEIC, we use the integrated luminosity in Ref. \cite{Accardi:2012qut}.  For the LHeC and EicC, we assume a luminosity of 1.0$\times 10^{33} \mathrm{cm}^{-2}\mathrm{s}^{-1}$ for $10^7$s of running. For RHIC, we assume a luminosity of $4.50\times 10^{29} \mathrm{cm}^{-2}\mathrm{s}^{-1}$ for $10^7$ s of running while for the LHC, we use the integrated luminosity from Chapter 10 of Ref. \cite{Citron:2018lsq}.
}
\label{tab:parameters}
\end{center}
\end{table} 

For $ep$ scattering, we follow the same approach that is in Ref. \cite{Lomnitz:2018juf}. 
\begin{eqnarray}
\sigma(ep\to eX^+n)&=&\int \frac{dk}{k}dQ^2\frac{d^2N_\gamma(k,Q^2)}{dkdQ^2}\notag\\
   &&\times\sigma_{\gamma^* p\to X^+n}(W,Q^2), 
\end{eqnarray}
where $k$ is the photon energy in the target rest frame, $W$ is the $\gamma^* p$ system center of mass-energy and $Q^2$ is the photon virtuality. The photon flux, $d^2N_\gamma(k,Q^2)/dkdQ^2$ is from Ref.~\cite{Budnev:1974de}, and  $\sigma_{\gamma^* p\to X^+n}(W,Q^2)$ is the $X^+$ photoproduction cross section.  
We model $\sigma_{\gamma^*p\to X^+n}(W,Q^2)$ following \cite{Adloff:1999kg}
\begin{eqnarray}
\sigma_{\gamma^*p\to X^+n}(W,Q^2)&=&\bigg(\frac{M_X^2}{M_X^2+Q^2}\bigg)^\eta\notag\\
  && \sigma_{\gamma p\to X^+n}(W,Q^2=0)f(M_X),
  \label{eq:eta}
\end{eqnarray}
where $f(M_X)$ is a relativistic Breit-Wigner function \cite{Soding:1965nh}, with mass $M_X$=1318 MeV and width $\Gamma=105 $ MeV 
for the $a_2^+(1320)$ and $M_X=4478$ MeV and $\Gamma=181$ MeV
for the $Z_c^+(4430)$ \cite{Tanabashi:2018oca},   We adopt the approach used in Ref. \cite{Klein:1999qj} to account for momentum-dependent broadening of the $\rho^0$ .  The choice of momentum broadening approach does not have a significant effect on the total cross-section \cite{Adamczyk:2017vfu}.

\begin{figure}[t]
	\centering
	\includegraphics[width=3 in]{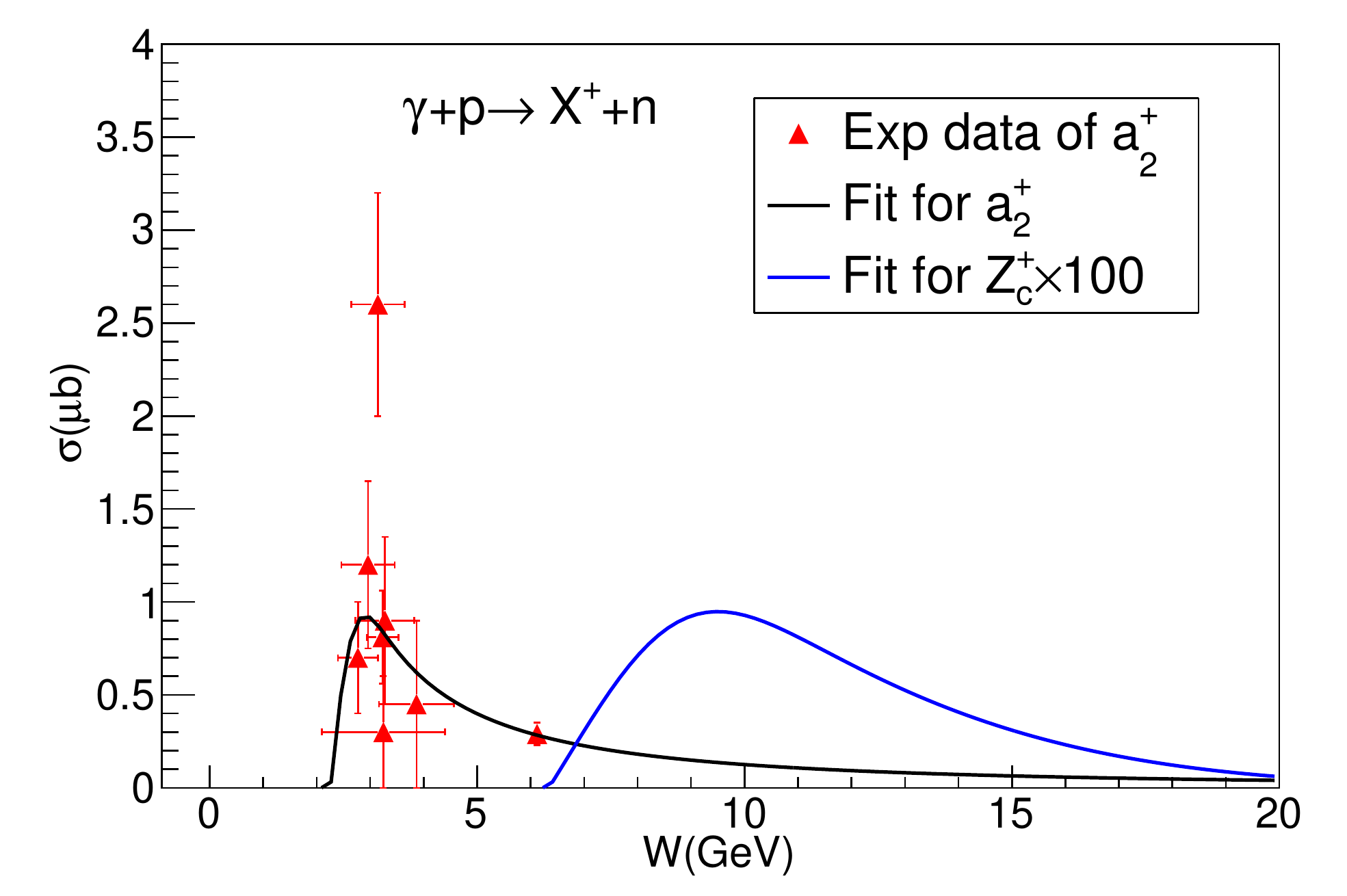}
	\caption{(Color online) Total cross section of $a_2^+(1320)$ and $Z_c^+(4430)$ as a function of $W$. The experimental data are taken from Refs.~\cite{Eisenberg:1969kk,Ballam:1971wq,Struczinski:1975ik,Condo:1993xa,Nozar:2008aa}.}
	\label{cs}
\end{figure} 

The variable $\eta=c_1+c_2(Q^2+M_X^2)$ controls the $Q^2$ dependence of the flux.  For vector mesons, $c_1$ and $c_2$ are determined from HERA data; they are unknown for the   $a_2^+(1320)$, so we use the  $c_1$ and $c_2$ determined for the $\rho$, while for the $Z_c^+(4430)$ we use the values from the $J/\psi$ \cite{Aaron:2009xp,Lomnitz:2018juf}.

We model $\sigma_{\gamma p\to a_2^+ n}(W,Q^2=0)$ with a  Regge-inspired parameterization of the fixed-target photoproduction data from Refs. \cite{Eisenberg:1969kk,Ballam:1971wq,Struczinski:1975ik,Condo:1993xa,Nozar:2008aa}, with four fitted parameters:
\begin{eqnarray}
&&\sigma_{\gamma p\to a_2^+(1320)n}(W)\approx5.42(W^2-m_p)^{-0.82}\notag\\&&-5.80\exp(-0.070(W^2-m_p^2)^2),
\label{a2_sigma}
\end{eqnarray}
where $W$ is the energy of the $\gamma p$ system in GeV,  $m_p$ is the proton mass and $\sigma$ is in $\mu$barn.  
Fig. \ref{cs} shows the data and fit.  Near the peak in the cross-section, at $W=2.9$ GeV,  the cross-section to produce an $a_2^+$ is about 7\% of that to photoproduce a $\rho^0$ ~\cite{Klein:1999qj}.  

We model $Z_c^+(4430)$  photoproduction following the Reggeon-model calculation in Ref. \cite{Galata:2011bi}.  Numerically, we use a 4-parameter fit to the $J^P=1^+$ curve in Ref. \cite{Galata:2011bi}:
\begin{eqnarray}
&&\sigma_{\gamma p\to Z_c^+(4430)n} (W)\approx0.0257\exp(-0.0094W^2)\notag\\&&-0.0317\exp(-0.00038(W^2-m_p^2)^2),
\label{Zc_sigma}
\end{eqnarray}
where $W$ is in GeV and $\sigma$ is in $\mu$barn.  This gives a result within a few percent of the published curve.  

\begin{figure*}[tbp]
	\centering
	\includegraphics[width=3in]{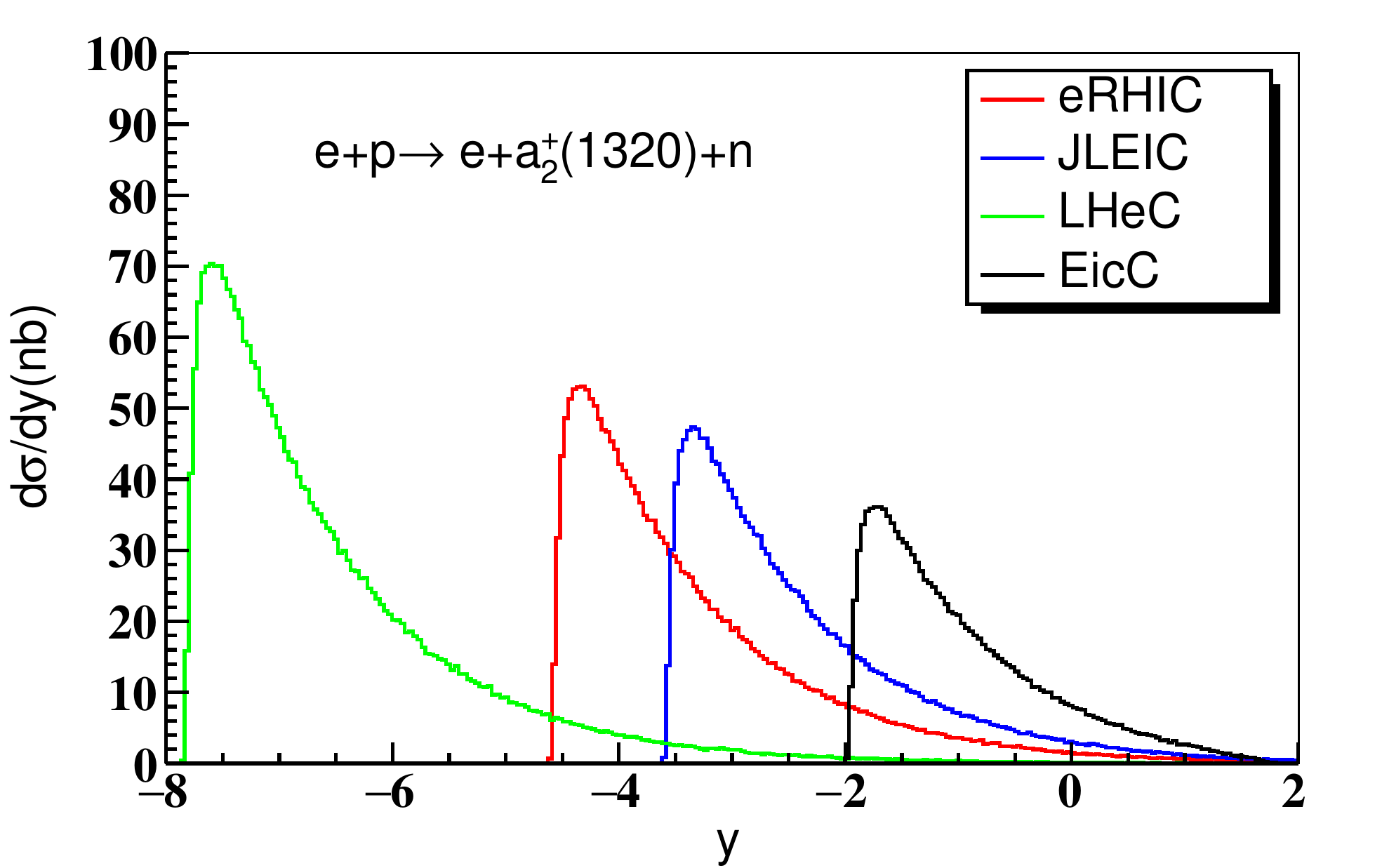}
	\includegraphics[width=3in]{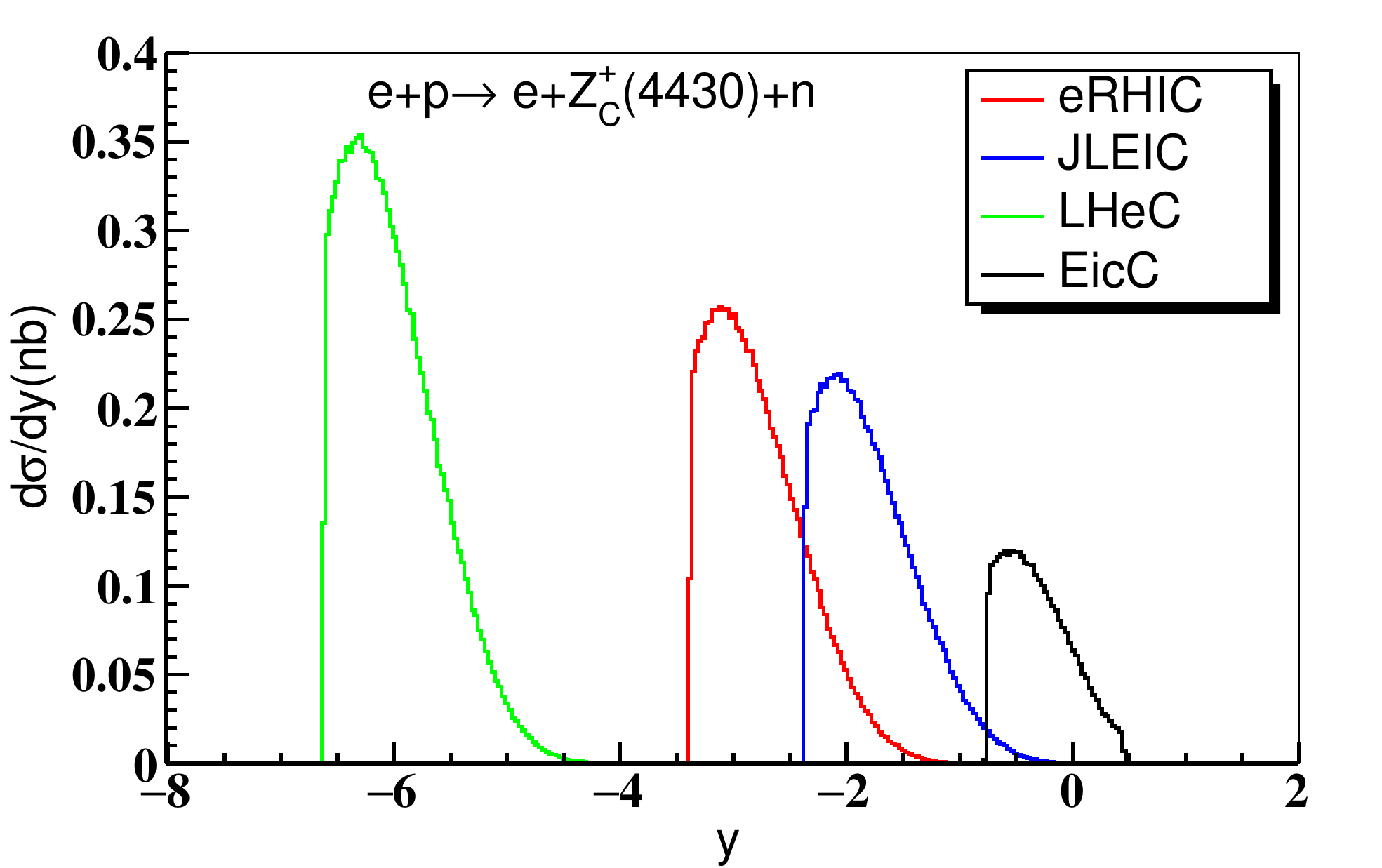}
	\caption{(Color online) Rapidity distributions of $a_2^+(1320)$ and $Z_c^+(4430)$ in $ep$ scattering at the planning EICs (0$<\mathrm{Q}^2<$1.0 $\mathrm{GeV}^2$). The electron is moving in the positive y direction.} 
	\label{ep_rap}
\end{figure*} 	

In UPCs, the cross section is \cite{Klein:2016yzr}
\begin{eqnarray}
\sigma(pA\to AX^+n)=\int dk\frac{dN_\gamma(k)}{dk}\sigma_{\gamma p\to X^+n}(W),
\end{eqnarray}
where $dN_\gamma(k)/dk$ is the photon flux of the ions~\cite{Bertulani:2005ru} and $\sigma_{\gamma p\to X^+n}(W)$ is the total cross section of photon-proton interaction.  Because the photon flux is proportional to the square of the ion charge, the photon flux of heavy ions dominates over  the photon flux of proton \cite{Bertulani:2005ru,Citron:2018lsq}.  Therefore, we  only consider the contributions of $\gamma p$ in photon flux emitted from heavy ions.

The $p_T$ distribution of the photoproduced particles is an important in background rejection, particularly for UPCs.  It includes contributions from the photon $p_T$ and  the Reggeon $p_T$, added in quadrature.  The former is modeled based on the proton form factor \cite{Klein:1999gv}, while, for the latter, we use a dipole form factor \cite{Klein:2003vd}.   

{\it Results}$-$The cross-sections and production rates for the six accelerators are shown in Tab. \ref{totalcs}.   For the EIC's, we present rates for two $Q^2$ ranges, $Q^2<1$ GeV$^2$, which we consider photoproduction, and 1 GeV$^2 < Q^2<5$ GeV$^2$, which we consider electroproduction; the rate for $Q^2>5$ GeV$^2$ is small.  As with vector mesons, the rate for electroproduction is a few percent of that for photoproduction \cite{Lomnitz:2018juf}.  The electroproduction rates are quite sensitive to the $\eta$ values in Eq. \ref{eq:eta}. 

The rapidity distributions  for photoproduction at an EIC are shown in Fig.~\ref{ep_rap}, where the incoming electron comes from negative rapidity.  The left plot is for the $a_2^+(1320)$, while the right is for the $Z_c^+(4430)$.  The shapes of these  distributions reflect the shape of cross sections presented in Fig. \ref{cs}.  The $a_2^+(1320)$ and $Z_c^+(4430)$ are produced in a relatively narrow rapidity regions because $\sigma(\gamma p\rightarrow X^+n)$ for both is largest near threshold, due to the Reggeon exchange mechanism.   This is very different from the Pomeron+ Reggeon exchange process which  spreads $\rho^0$ over a wide rapidity range \cite{Lomnitz:2018juf}.

The pseudorapidity distribution of the final state pions and leptons is somewhat broader than the rapidity distribution of their photoproduced produced parents.   Reggeon exchange reactions can be easier to study at moderate energy EICs; the LHeC would be a difficult environment for these studies.  Detailed observation will require detectors with good capabilities in the forward region.    Alternately, an EIC could be run with a lowered proton/ion beam energy, to shift production into a mid-rapidity detector.  Still,   Other mesons, with masses between the $a_2^+(1320)$ and the $Z_c^+$ should be produced at intermediate rapidities between the two.  

\begin{figure*}[htbp]
	\centering
	\includegraphics[width=2.3in]{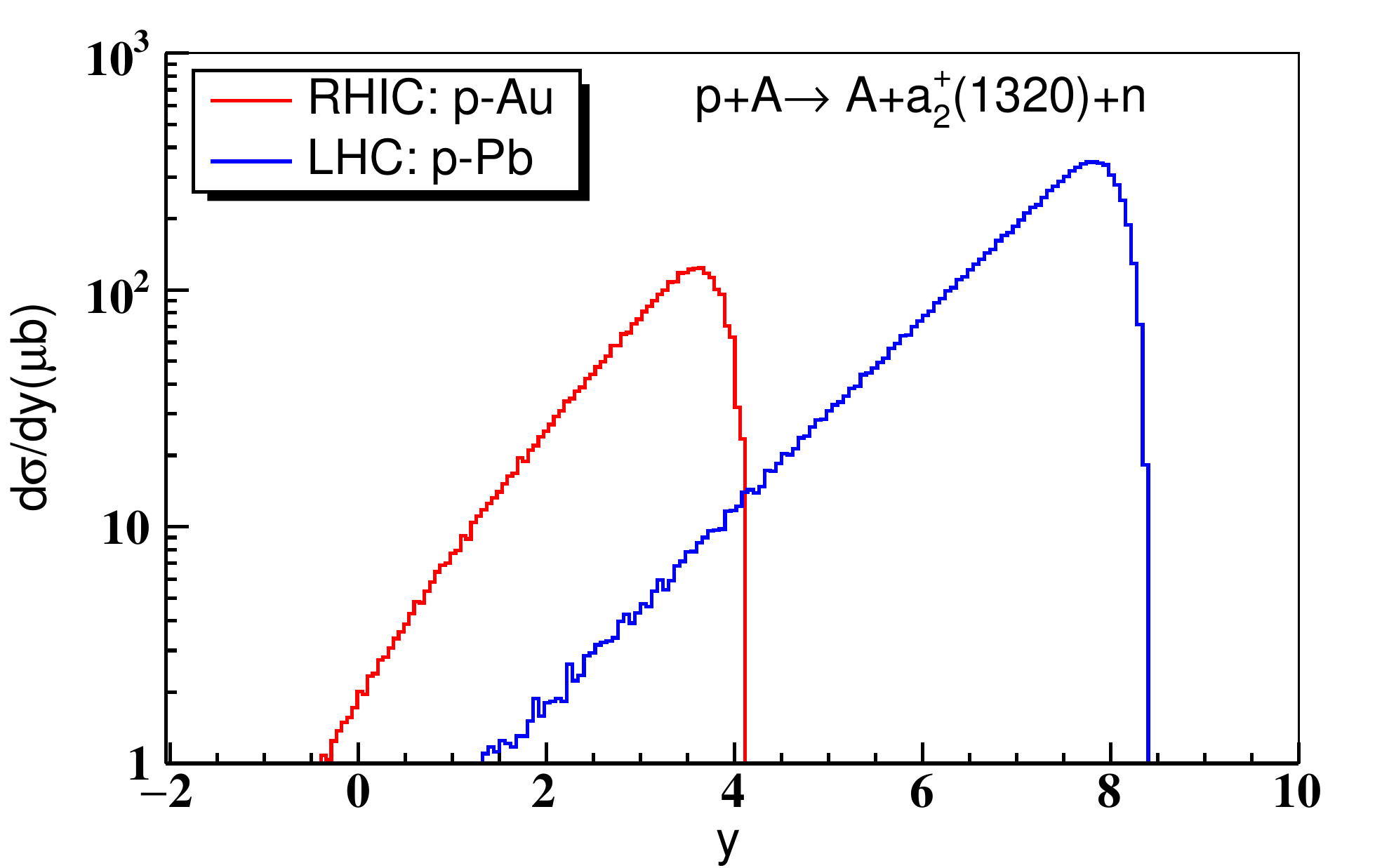}
	\includegraphics[width=2.3in]{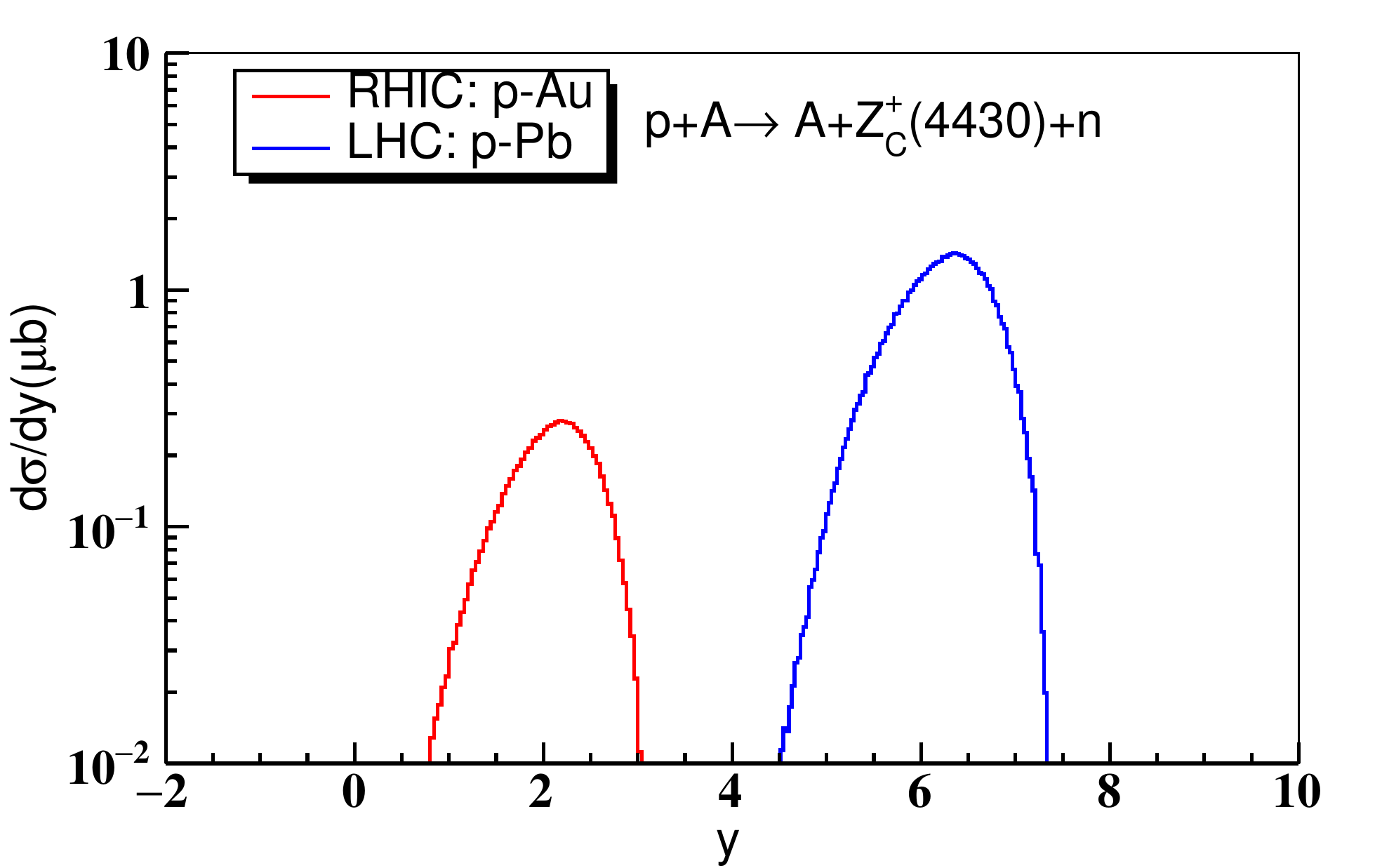}
   \includegraphics[width=2.3in]{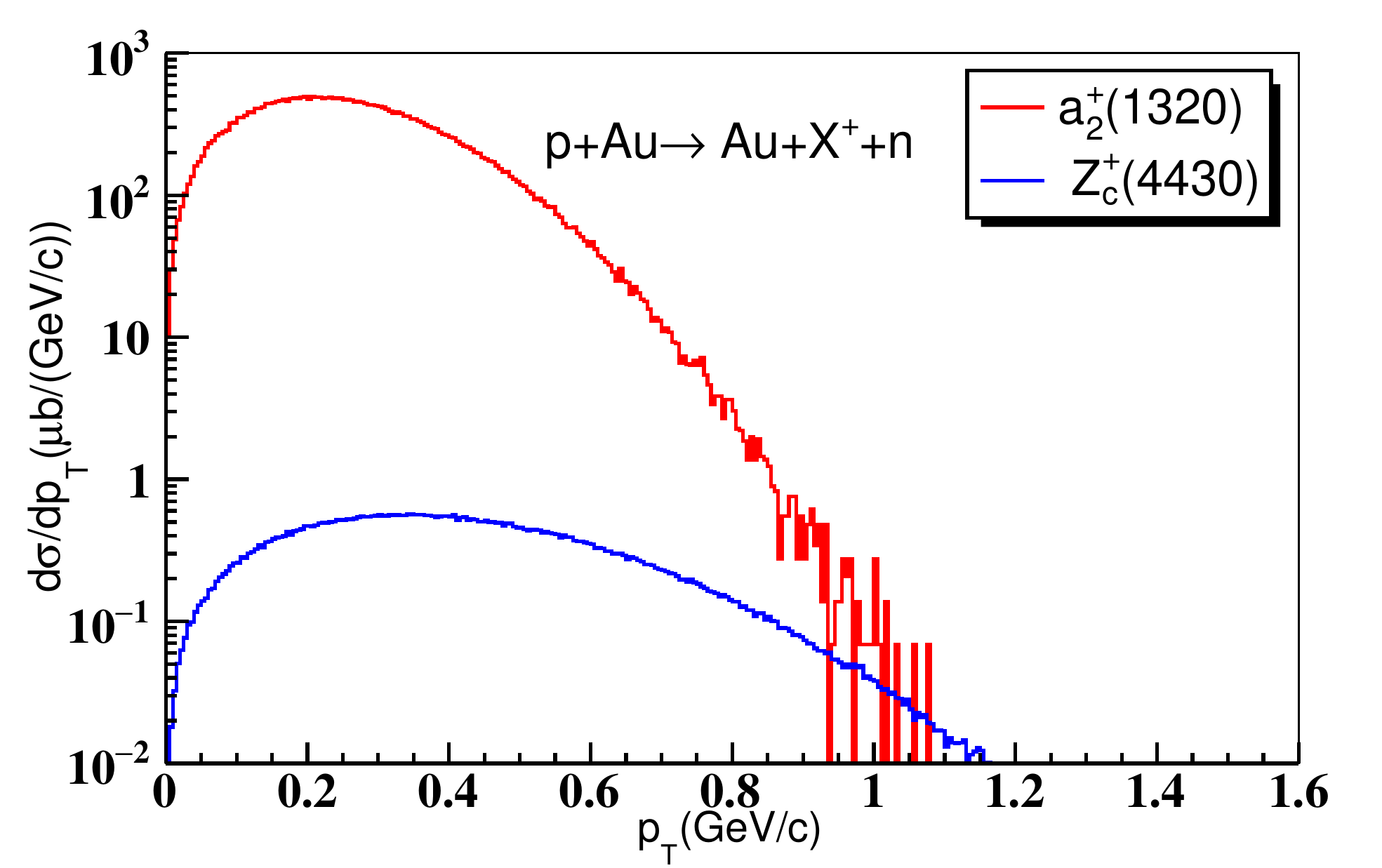}
	\caption{(Color online) Rapidity distributions of $a_2^+(1320)$ and $Z_c^+(4430)$ in $pA$ collisions at RHIC and the LHC and transverse momentum distributions of $a_2^+(1320)$ and $Z_c^+(4430)$ in $pAu$ UPCs at RHIC. The proton is moving in the positive $y$ direction.} 
	\label{pA_rap}
\end{figure*}

Pseudorapidity acceptance is also an issue at RHIC and the LHC, where central detectors typically cover $|\eta|<1$ up to $|\eta|<2.4$.  
Fig.~\ref{pA_rap} (left) shows rapidity distributions for the $a_2^+(1320)$ in $pAu$ UPCs at RHIC and $pPb$ UPCs at the  LHC, 
while Fig.~\ref{pA_rap} (middle) shows rapidity distributions for the $Z_c^+(4430)$ in $pAu$ UPCs at RHIC and $pPb$ UPCs at the  LHC. 
Because the lower beam energy leads to more central production,   At the LHC, the LHCb experiment covers $2 < \eta <5$ \cite{McNulty:2019hld}, so should have good acceptance for both the $a_2^+(1320)$ and the $Z_c^+(4430)$.  At RHIC, a detector with moderately forward acceptance could also study the $a_2^+(1320)$.

Fig.~\ref{pA_rap} (right) shows transverse momentum distributions for $a_2^+(1320)$ and $Z_c^+(4430)$ in $pAu$ UPCs at RHIC. These spectra are similar to those for vector mesons with similar masses, and should be similarly easy to detect.   

Even after accounting for detector acceptance and the branching ratios,  the rates for the $a_2^+ (1320)$ should be very high at the U. S. and Chinese EICs and at RHIC, enough for detailed studies of the energy dependence and (for the EIC) $Q^2$ dependence  of the production cross-section.   Many other light-quark hadrons should also be accessible.  The rates for the $Z_c^+(4430)$ are lower, but still high enough for studies of the cross-section to confirm (or refute) it's nature as a tetraquark and shed light on its spin.  

\begin{table*}[t]
\begin{tabular}{|c |c |c | c| c| c|c|c|c|c|c|}
			\hline
			\hline
&  \multicolumn{4}{c|}{$\sigma$ ($0 <\mathrm{Q }^2< 1.0 \mathrm{ GeV}^2$ )} &\multicolumn{4}{c|} 
 {$\sigma$  ($1.0\mathrm{ GeV}^2<\mathrm{ Q}^2 < 5.0 \mathrm{ GeV}^2$ ) }& \multicolumn{2}{c|}{$\sigma$ ($\mathrm{Q }^2=0.0$ ) } \\	
		& eRHIC &JLEIC& LHeC& EicC &  eRHIC &  JLEIC& LHeC& EicC &RHIC   & LHC  \\
		\hline
$a_2^+(1320)$ & 79 nb    & 69 nb    & 106 nb    & 47 nb  & 0.51 nb  & 0.50 nb  &  0.52 nb   &  0.40 nb  & 0.17 mb   & 0.56 mb    \\
$Z_c^+(4430)$ & 0.26 nb    & 0.22 nb    & 0.36 nb   &  0.094 nb & 12.17 pb  & 11.75 pb   &  12.27 pb  &  6.80 pb  &  0.32 $\mu$b  &   1.76 $\mu$b \\
			\hline
& \multicolumn{4}{c|}{Event  ($0 <\mathrm{Q }^2< 1.0 \mathrm{ GeV}^2$ )} &\multicolumn{4}{|c|} 
{Event   ($1.0\mathrm{ GeV}^2<\mathrm{ Q}^2 < 5.0 \mathrm{ GeV}^2$ ) }& \multicolumn{2}{c|}{Events  ($\mathrm{Q }^2=0.0$ ) } \\	
& eRHIC &JLEIC& LHeC& EicC &  eRHIC &  JLEIC& LHeC& EicC &RHIC   &  LHC  \\
\hline
$a_2^+(1320)$ & 0.79 B    & 0.69 B    & 1.06  B    & 0.47 B  & 5.1 M  & 5.0 M  &  5.2 M   &  4.0 M  & 0.78 B   & 1.12 B \\
$Z_c^+(4430)$ & 2.6 M    & 2.2 M    &3.6 M  &  0.94 M & 0.12 M   & 0.12 M   & 0.12 M &  68.0 K &  1.4 M  & 3.52 M \\
\hline 			
\hline
\end{tabular}
\caption{Total cross section and event rates for $a_2^+(1320) $ and $Z_c^+(4430)$ photoproduction in proposed $ep$ scattering and proton-nucleus UPCs. Here, $B=10^9$, $M=10^6$ and $K=10^3$. The integrated luminosities are taken from Table.~\ref{tab:parameters}}
\label{totalcs}
\end{table*} 

Studies with $pA$ UPCs at RHIC could quickly lead to relatively precise measurements of the $a_2^+(1320)$ photoproduction cross-section and establish a benchmark for Reggeon-exchange photoproduction studies.  Other lighter mesons should be accessible using the same approach.  With enhanced forward detectors (or lower-energy collisions), the $Z_c^+(4430)$ and other $c\overline c$ based exotic should also be within experimental reach.  

{\it Conclusions}$-$We have calculated the photoproduction cross-sections, $d\sigma/dy$ and event rates for the $a_2^+(1320)$ and $Z_c^+(4430)$
at four EICs and two hadron accelerators.    In all six cases, the total $a_2^+(1320)$ production rates are high - of order 1 billion events per year.   The electroproduction rates are also high - millions of events per year.  With a wide-acceptance detector, it will be possible to study photoproduction over a wide range of $W$ and $Q^2$, and also study rare decays and the spin-structure of the production.  At RHIC, it will also be possible to study photoproduction with a polarized proton target. 

The photoproduction rates for the $Z_c^+(4430)$ are lower - of order 1 million events per year - but are still adequate for moderate precision studies, even after accounting for the lower branching ratios. Because the system is very clean, photoproduction will be a good place to search for different decay modes.  Detection of the outgoing neutron in zero degree calorimeters, and, for the EIC, the scattered electron, will allow for compete event reconstruction.  At an EIC, it may be possible to search for final states with missing particles using missing-mass techniques; this would allow for branching-ratio independent measurements of production cross-sections.

The major experimental challenge is that production is concentrated near the $\gamma p$ energy threshold, so, at high energies, production is in the far forward region.   However, they can be seen fairly readily in UPCs at RHIC, and at lower energy EICs.

Although we have focused on proton targets, Reggeon-exchange  reactions should also occur for nuclear targets.   Because the final nuclear state is altered, coherence may be lost, so the cross-sections will be lower.   Despite the low cross-sections, Reggeons may be an interesting probe of sea quarks in heavy nuclei.  Unlike Pomerons, which are mostly gluons, Reggeons represent meson exchange, so are composed mostly of quarks.   They may thus be a way to probe the distribution of sea quarks and antiquarks in nuclei \cite{Adamczyk:2017vfu}.

This work is supported by U.S. Department of Energy, Office of Science, Office of Nuclear Physics, under contract number
DE-AC02-05CH11231 and Chinese Academy of Sciences Scholarship Program.  We thank Nu Xu for useful discussions about EicC, and SK thanks Mark Strikman for useful conversations.

\end{document}